\documentstyle[12pt,epsf]{article} 
\newcommand{\be}{\begin{equation}} 
\newcommand{\en}{\end{equation}}
\newcommand{\bea}{\begin{eqnarray}}
\newcommand{\ena}{\end{eqnarray}}

\newcommand{\hbo}{\hbox to 1 true cm {\hfill } } 
\newcommand{\tr}{\hbox{tr}}
\newcommand{\Tr}{\hbox{Tr}}

\def\dslash{\partial\kern-.6em\slash}
\def\kslash{k\kern-.5em\slash}
\def\pslash{p\kern-.4em\slash}
\def\Dslash{D\kern-.6em\slash}
\def\Vslash{V\kern-.7em\slash}
\def\vslash{v\kern-.5em\slash}
\def\rslash{r\kern-.5em\slash}
\def\qslash{q\kern-.5em\slash}

\begin{document} 
\vglue 1truecm
  
\vbox{ UNITU-THEP-4/97
\hfill May 28, 1997
}
  
\vfil
\centerline{\large\bf A New State Of Hadronic Matter At High Density} 
  
\bigskip
\centerline{Kurt Langfeld$^a$, Hugo Reinhardt$^a$ and Mannque Rho$^b$ }
\vspace{1 true cm} 
\centerline{ $^a$ Institut f\"ur Theoretische Physik, Universit\"at 
   T\"ubingen }
\centerline{D--72076 T\"ubingen, Germany}
\bigskip
\centerline{ $^b$ Service de Physique Th\'eorique, C.E. Saclay, } 
\centerline{ F-91191 Gif-sur-Yvette Cedex, France } 
\centerline{and}
\centerline {Departament de Fisica Te\`orica, Universitat de Val\`encia}
\centerline{E-46100 Burjassot (Val\`encia), Spain} 
  
\vfil
\begin{abstract}
We propose in this article that if the chemical potential exceeds a 
critical value in dense hadronic medium, a first-order phase transition 
to a new state of matter with  
Lorentz symmetry {\it spontaneously broken} (in addition to the explicit 
breaking) takes place.  As a consequence, light vector mesons 
get excited as ``almost'' Goldstone bosons. 
Since the light vector mesons dominantly couple to photons, the presence 
of these new vector mesons could lead to an enhancement in the 
dilepton production from dense medium at an invariant mass lower than 
the free-space vector-meson 
mass. We provide a low-energy quark model which demonstrates 
that the above scenario is a generic case for quark 
theories with a strong interaction in the vector channel. We discuss
possible relevance of this phase to the phenomenon of
the enhanced dilepton production at low invariant masses
in relativistic heavy-ion collisions.

\end{abstract}

\vfil
\hrule width 5truecm
\vskip .2truecm
\begin{quote} 
\end{quote}
\eject

\section{ Introduction }
\vskip 0.3cm 

Recently, the CERES and  HELIOS collaborations reported the exciting 
observation~\cite{ceres,helios} that 
the lepton pair production is enhanced in S--Au collisions in the 
invariant mass range of $300 \ldots 600 \, $MeV compared with the collisions 
p--Be and p--Au. 
This observation provides an important clue as to
what happens to hadronic matter when it is compressed to a high density.
Unlike the case of temperature, lattice calculations are not yet in a
position to provide information on the effect of density on QCD vacuum
and hence practically nothing is understood of density-driven QCD phase 
transitions. Random matrix studies show indeed that the effect of a 
chemical potential can be exceedingly subtle from the point of view of 
QCD \cite{nowak}.
In the paucity of any first-principle guidance, there is a wide range
of theoretical ideas to explore.
It will ultimately be up to experiments to weed out wrong ideas and
to guide us towards a viable scenario.
\vskip 0.3cm

From a detailed study of the collisions with the help 
of covariant transport equations, it became clear that the dilepton 
yield of the collisions p--A in the above mass range can be well 
understood by resorting only to the decays of the $\eta $, $\rho $, $\omega $ 
and $\phi $-mesons~\cite{cass95}. The fact that a large pion density is 
produced by the collision and the experimental observation that the 
dilepton enhancement sets in at roughly twice the pion mass have led to the 
conjecture that the enhancement is due to $\pi ^+ \pi ^-$--annihilation 
processes~\cite{ceres}. The quantitative analysis, however, showed that 
the experimental data cannot be explained using ``free'' meson masses and 
form factors~\cite{cass95,wam96}. A change of the state of matter or 
at least a change of the meson properties in medium seem to be required. 
Several proposals~\cite{schuck,wam96,son96,hat92,rho96,rho91} 
have been made, most of which focusing on the role 
of the $\rho $ vector meson. This meson is of particular importance 
for the dilepton enhancement effect, since the $\rho $-meson directly 
couples to the 
photons. The coupling of the $\rho $-meson to two pion states and 
the change of the pion propagation in medium generically results 
in a broadening of the $\rho $-meson peak~\cite{wam96,son96}. 
The observed increase in the dilepton production rate is compatible 
within the error bars of the present data. On the other hand, QCD 
sum rules~\cite{hat92} predict a decreasing $\rho $-meson mass for 
increasing matter density. Whereas the sum rule approach is 
restricted to small densities, a decrease of the $\rho $-mass 
as function of the matter density should hold due to the onset 
of the chiral phase transition~\cite{rho96}. These considerations 
supplemented with further support from the results from 
the Skyrme model can be summarized in the scaling in medium of the hadron
``quasi-particle'' masses known as BR-scaling~\cite{rho91}. Including a 
$\rho $-meson mass shift to smaller values in the calculation of 
the relativistic transport theory, a good agreement of the theoretical 
predictions with the observed dilepton spectra in the S--Au collision 
is achieved~\cite{ko96}. The two approaches, one based on many-body 
correlations starting from strongly coupled hadrons whose properties
are defined in matter-free space~\cite{wam96,son96} (with broadening
widths) and 
the other based on the notion
of both bosonic and fermionic {\it quasi-particles} with parameters defined 
in a medium background field~\cite{SBMR}, somewhat contradictory to each
other
though they may appear to be, are probably related to each other
when applied to the dilepton phenomena in question. Whether or not the two
ways of looking at dense matter can be mapped to each other for other
physical observables is not clear.
%
\vskip 0.3cm

In this paper, based on a rather generic argument,
we propose a novel state of hadronic matter which is 
unstable in the (zero-density) vacuum, but which forms the state 
with the lowest 
energy density, if matter is present. In addition to the explicit breaking 
of Lorentz invariance due to a finite baryonic density, Lorentz 
symmetry is also {\it spontaneously broken} in this new matter state.
This provides a mechanism for exciting low-mass vector mesons 
as ``almost'' Goldstone 
bosons. We will discuss the properties of such vector mesons in some detail. 
Our considerations are based solely on the realization of symmetries. 
This model-independent argument will be given a support by  
an effective low-energy quark model which will illustrate that the new 
matter state is generically present in theories with vector-type 
quark interactions as it is the case for QCD.
 
\vskip 0.3cm
 
\section{ Induced Spontaneous Symmetry Breaking}
\vskip 0.3cm 

In matter-free space, the QCD vacuum is 
characterized by a non-vanishing value of the quark condensate and 
zero baryonic density, i.e. 
\be 
\langle \bar{q} q \rangle \; \not= \; 0, \; \hbo 
3 \, \rho_B = \langle \bar{q} \gamma _0 q \rangle \; = \; 0 \; 
\hbox to 4 true cm { \hfill State (V)}. 
\label{eq:1} 
\en
The non-zero condensate implies that chiral symmetry of QCD 
is spontaneously broken (apart from a small explicit breaking through 
current quark masses). This particular realization 
of chiral symmetry allows the interpretation of the light pseudo-scalar mesons 
as Goldstone bosons~\cite{nam61} and provides a model-independent 
explanation of the particular role of the pion in the meson mass spectrum. 
\vskip 0.3cm

The key observation in this paper is that there is a second state 
which is not realized in the vacuum but appears as a meta-stable state 
having a higher energy density than the vacuum. 
This additional state 
is characterized by a vanishing (scalar) quark  condensate, and a 
vanishing baryon density (represented in terms of quark fields) 
$\langle \bar{q} \gamma _0 q \rangle $, i.e. 
\be 
\rho _B, \, \langle \bar{q} q \rangle \; = \; 0, \; \hbo 
\zeta ^2 := \langle \bar{q} \gamma _\mu q \; \bar{q} \gamma ^\mu q \rangle 
\; \not= \; 0 
\hbox to 4cm { \hfill State (II)  }, 
\label{eq:3} 
\en
and the symmetry is in a Kosterlitz-Thouless type of 
realization~\cite{wi78}. 
\vskip 0.3cm

Our crucial point is that when a small explicit breaking of 
Lorentz symmetry is introduced via the chemical potential,
two important things happen.  First, at certain chemical potential $\mu_c$, 
the state (II) becomes energetically favored and a phase transition 
takes place from the state (V) to 
the state (II). Second, the presence of matter
 selects a particular Lorentz frame, since it favors the zero component 
of the vector current.  The matter state with the lowest energy density 
is then described by 
\be 
\langle \bar{q} q \rangle \; = \; 0, \; \hbo 
3 \, \rho_B = \langle \bar{q} \gamma _0 q \rangle \; \not= \; 0 
\hbox to 4cm { \hfill State (M) }. 
\label{eq:2} 
\en
In addition to the contribution due to the chemical potential, the baryonic 
density acquires a large contribution from the dynamics of the 
theory. Since the chemical potential $\mu $ must exceed a critical 
value $\mu_c$ to generate this dynamical contribution, we shall refer to 
this scenario as induced spontaneous symmetry breaking (ISSB). 
\vskip 0.3cm

One might object at this point 
that the ISSB scenario is in contradiction to the Vafa-Witten 
theorem~\cite{vaf84}, which states that vector symmetries cannot be 
spontaneously broken in QCD. In fact, what the theorem is telling us is that 
the correlation 
function in the vector channel has, for a given gluon configuration, an upper 
bound provided by an exponentially decreasing function of distance. 
Since the QCD weight 
as employed by averaging over all gluon configurations is positive, the full 
correlation function has the same upper bound. This would rule out a massless 
vector state. 
In our case, the state (M) becomes the ground state for $\mu > \mu_c$, with
the vector particle becoming light, but not 
massless due to the additional {\it explicit} breaking of Lorentz symmetry. 
Therefore the correlation function will be decreasing with an 
exponential slope. Thus the ISSB scenario does not contradict the Vafa-Witten theorem\footnote{Furthermore, the Vafa-Witten theorem is proven in
Euclidean space. In the presence of a chemical potential 
(and a gluonic background field), the determinant
develops a phase, upsetting the positivity of the measure required for the
proof. We would like to thank Maciek Nowak for reminding us of this
caveat.}. 
\vskip 0.3cm

Let us discuss the consequences of the ISSB scenario. Assume that 
the matter phase is realized in the state (M) and that the small explicit 
breaking via the chemical potential $\mu > \mu _c$ can  be taken into 
account perturbatively. 
Due to the presence of the condensate $\langle \bar{q} \gamma _0 q \rangle$, 
the symmetry with respect to Lorentz boost transformation, 
$\Lambda ^{0 }_{\phantom{0} \nu } = 
\exp \{ \omega  \}^{0 }_{\phantom{0 } \nu }$, 
\be 
q(x) \rightarrow q^{\prime } (x^\prime ) \; = \; S(\Lambda) q(x) \; , 
\hbo 
S(\Lambda ) \; = \; \exp \{ - \frac{i}{4} \omega _{\mu \nu } 
\sigma ^{\mu \nu } \} \; , 
\label{eq:4} 
\en 
is spontaneously broken (in addition to the small explicit breaking 
induced by $\mu $). What is the corresponding Goldstone boson? In 
order to answer this question, we resort to standard techniques which 
were developed in the context of the spontaneous breakdown of chiral 
symmetry~\cite{go84}. For this purpose, first note that the 
quark propagator of the state (M) satisfying the Dyson-Schwinger equation 
possesses the general structure 
\be 
s(k) \; = \; \frac{ Z(k_0, \vec{k} ) }{ 
\kslash \; - \; V_0 (k_0, \vec{k} ) \gamma _0 \, - \Sigma (k_0, \vec{k}) 
} \; . 
\label{eq:5} 
\en 
In the low-energy regime (where the momentum transfer $k$ is much smaller 
than the typical gluonic energy scale), one expects that the quark theory 
can be approximated by a Nambu-Jona-Lasinio type effective model in which case
the momentum and energy dependence of the functions $Z$, $V_0$ and 
$\Sigma $ in (\ref{eq:5}) can be neglected~\cite{la96}. For simplicity, 
we will make this assumption in the following. Furthermore note 
that the transformed propagator, i.e. 
\be 
S(\Lambda ) s(k ^\prime ) S^{-1}(\Lambda ) \; , 
\label{eq:6} 
\en 
also satisfies the Dyson-Schwinger equation, since the latter equation is 
manifestly Lorentz covariant for $\mu =0$. Considering infinitesimal 
Lorentz boost transformations  
induced by $\omega _{0i}$, one concludes that the vertex function 
\be 
P^V \; \propto \; V_0 \, [\gamma _0 , \sigma _{0i} ] \; \propto \; 
V_0 \, \gamma _i 
\label{eq:7} 
\en 
satisfies a Bethe-Salpeter equation with zero mass. 
The important finding is that,  at least at 
sufficiently small energies, the corresponding particle is of pure 
vector type. The emergence of the light vector particle is a 
consequence of the spontaneous breaking of the Lorentz group down 
to the non-relativistic rotational group SO(3) and is analogous to the 
emergence of scalar Goldstone bosons for a spontaneously broken 
internal symmetry group. Since in the present case the broken 
symmetry is a space-time symmetry, the corresponding massless 
particles are (non-relativistic) spin one excitations 
(for more details see section 3). 
\vskip 0.3cm 

If one wants to relax the restriction to the low-energy regime, 
one might resort to  the full vertex function, which is provided by the 
Noether charge density generated by the Lorentz boost 
(\ref{eq:4}), i.e. 
\be 
Q^{(L)} _i(x) \; = \; (x^0 T^{0i}(x) - x^i T^{00}(x)) \; + \; 
\bar{q}(x) \gamma _i q(x) \;  , 
\label{eq:8} 
\en 
where $T^{\mu \nu }$ is the energy-momentum tensor written in terms of 
quark fields. Here, the first term represents the ``orbital'' part while 
the second term gives rise to the ``spin'' part of the vertex function. 
\vskip 0.3cm

The state (M) in (\ref{eq:2}) exists only for finite values of the 
chemical potential $\mu $, which explicitly breaks Lorentz invariance. 
This explicit breaking has the consequences that the vertex function 
$P^V$ in (\ref{eq:7}) acquires additional parts and that the 
Goldstone vector particle gets a small mass. 
To demonstrate this, we use the variational 
approach of~\cite{la90}, which is a kind of relativistic 
RPA approach -- and  a convenient one if the light mesons are 
treated as Goldstone bosons. 
As a trial state for the variational approach, we allow 
the vertex function $P^V$ to gain additional vector parts at finite values 
of $\mu $ . The vector field operator is chosen in terms of the 
quark operators as 
\be 
{\cal V} _i (t) \; = \; \kappa _L \, \int d^3x \; Q^{(L)}_i(x) 
\; + \; \kappa _V \, \int d^3x \; \bar{q}(x) \gamma _i q(x) \; , 
\label{eq:9} 
\en 
where $\kappa _{L/V}$ are variational constants. 
Using the techniques of ~\cite{la90}, it is straightforward to relate 
the mass of the vector Goldstone boson $m_V$ to the explicit breaking 
via the chemical potential $\mu $. We find 
\be 
m_V^2 f_V ^2 \; = \;  2 \mu \langle \bar{q} \gamma _0 q \rangle \; , 
\label{eq:10} 
\en 
where $f_V$ is a decay constant defined by 
\be 
\Big\langle \Omega \Big\vert Q^{(L)}_i (x=0) \Big\vert {\cal V}_k 
(m_V,\vec{p}=0) \Big\rangle \; = \; i f_V \, m_V \; \delta _{ik} \; . 
\label{eq:11} 
\en 
Here $\vert {\cal V} (p) \rangle $ denotes the state of the 
Goldstone vector with four momentum $p$, and $\vert \Omega \rangle $ 
is the matter ground state. Equation (\ref{eq:10}) is nothing but the 
analog of the Gell-Mann-Oakes-Renner relation~\cite{ch84}. 
The decay constant $f_V$ describes the decay of the Goldstone vector 
into photons. A calculation of this quantity in a simple model will
be presented later. Since the electro-magnetic U(1) gauge invariance is 
broken by the presence of matter, the coupling of quarks and 
photons is non-minimal, but acquires additional pieces proportional 
to the ''angular momentum'' density (the first term in eq.~(\ref{eq:8})). 
\vskip 0.3cm 

The local minimum of the effective potential at $\langle \bar{q} q 
\rangle =0 $ and $\langle \bar{q} \gamma _0 q \rangle \not= 0 $ 
(i.e. the state (M)) becomes the global minimum, i.e. the ground state, 
when $\mu $ exceeds a critical value. In this new state, the 
condensate $\langle \bar{q} \gamma _0 q \rangle $ does not scale with 
$\mu $ anymore, but acquires a strong contribution from the 
interaction. As shown above, this mechanism results in a light 
iso-scalar vector meson. The occurrence of the light vector meson 
is obviously accompanied by the spontaneous generation of baryon density 
$\langle \bar{q} \gamma _0 q \rangle $. On the other hand, the 
$U(1)_V$ vector symmetry is conserved in our approach implying that 
baryon number is conserved. Both statements above do not contradict 
each other, since we have assumed an infinite system at finite baryon 
density. In practical applications a finite baryon number 
localized in space is the interesting case. 
In this case, we cannot assume a homogeneous phase. Rather we expect 
that the gradients discarded in the above description will lead to 
a domain structure, where each domain is characterized by 
a different value of the order parameter $\langle \bar{q} \gamma _0 q 
\rangle $. 
\vskip 0.3cm

In order to roughly estimate the order of magnitude of the parameters 
involved, we assume that the phase transition of the vacuum 
(\ref{eq:1}) to the state (M) (\ref{eq:2}) occurs at a Fermi momentum 
$k_f \approx M_c$, where $M_c \approx 300 \, $MeV is the constituent 
quark mass. This is the standard value for $k_f$, where one expects the 
chiral phase transition to occur. The estimate (based on a constituent 
quark model) of the corresponding chemical potential is 
therefore $\mu \approx \sqrt{ M_c^2+k_f^2} \approx \sqrt{2} M_c$. 
Note that this value for $\mu $ is of the same order of the magnitude 
as the generic energy scale of the phase transition implying that 
corrections to (\ref{eq:10}) might become important. 

For a first guess, we further use the relation $\rho _B \approx 
2k_f^3/3\pi^2$ of the constituent quark model. 
Combining these rough estimates, we find $m_V f_V \sim (260 \, \hbox{MeV} )^2 $
from (\ref{eq:10}). If the mass of the ``almost'' Goldstone vector 
bosons is small, 
the resonance will be broad, since the coupling to the photons $f_V$ becomes 
large. 
\vskip 0.3cm

\section{ Model Calculation }
\vskip 0.3cm 

The results of the previous section are model-independent, 
 relying solely on a specific phase structure of the field theory. 
In this section, we illustrate with the help of a simple model 
that the particular phase structure required by the ISSB scenario 
is actually a generic case for effective 
quark models with a strong vector-current interaction. 
\vskip 0.3cm

For this purpose, we study an effective low energy quark 
model~\cite{nam61} defined at finite chemical potential $\mu $, 
by the generating functional for Euclidean Green's 
functions (see e.g.~\cite{eb86}) 
\bea 
Z[s,j_\mu] &=& \int {\cal D} q \, {\cal D} \bar{q} \, {\cal D} \sigma \, 
{\cal D} \pi \; {\cal D} V_\mu \; e^{ 
\int d^4x \; [L + s(x) \sigma (x) + j_\mu (x) V^\mu (x) ]} \; , 
\label{eq:12} \\ 
L &=& \bar{q}(x) \Big( i \dslash -  \sigma (x) + i \gamma _5 \pi (x) 
+ i V_\mu (x) \gamma ^\mu \Big) q(x) 
\label{eq:13} \\ 
&-& \frac{N}{2} g_s \left[(\sigma (x)-m)^2 + \pi ^2(x) \right] 
\nonumber \\ 
&-& \frac{N}{2} \left\{ V_\mu (x) \left( - \partial ^2 \delta _{\mu \nu } 
+ \partial _\mu \partial _\nu \right) V_\nu (x) 
\; + \; m^2_v \left[ (V_0(x) - \mu )^2 + V_k^2(x) \right] \right\} \; , 
\nonumber 
\ena 
where $N$ is the number of colors (the color index of the quark fields is 
not shown) and $m$ is the current quark mass. 
Our definitions of the Euclidean space can be found in appendix A. 
In particular, we have defined the square of an Euclidean vector by 
$\partial ^2:= \partial _\mu \partial _\mu = - \partial _\mu \partial ^\mu $. 
For simplicity we consider the case of one light-quark flavor. 
Generalizations to the iso-spin $I=1/2$ will be discussed in the next 
section. 
\vskip 0.3cm

Let us look at the model (\ref{eq:12}) in Minkowski space. 
For this purpose, we integrate out the meson fields $\sigma $, $\pi $ 
and $V_\mu $ and perform the analytic continuation of the 
Euclidean quark theory to Minkowski space. Our conventions can be 
found in appendix A. Integrating out the vector fields $V_\mu $ yields 
a non-local current-current interaction. If we neglect the momentum 
transfer in this interaction, the low energy quark theory is of NJL-type, 
i.e. 
\bea 
L_M &=& \bar{q}_M (x)( i\dslash - m - \mu \gamma ^0_{(M)} ) q_M (x) 
\label{eq:12b} \\ 
&+& \frac{1}{2Ng_s} \left[ (\bar{q}_M(x) q_M(x))^2 \, - \, 
(\bar{q}_M(x) \gamma _5 q_M(x))^2  \right] 
\nonumber \\ 
&+& \frac{1}{2N m_v^2} \, 
\bar{q}_M(x) \gamma ^{(M)} _\mu q_M(x) \, \bar{q}_M(x) 
\gamma ^{(M) \, \mu } q_M(x) \; . 
\nonumber 
\ena 
Note that the vector current interaction contributes with a plus sign 
to the Lagrangian $L_M$. We call this an attractive interaction in the 
vector channel. It will turn out that this sign is crucial for the 
ISSB scenario. We should stress that there is nothing to indicate that
such an interaction is incompatible with light-quark hadron phenomenology.
Now if we compare (\ref{eq:12b}) with (\ref{eq:13}), 
we find that the parameter $\mu $ in (\ref{eq:13}) can be interpreted 
as the chemical potential. Any 
non-vanishing expectation value of $V_0$ obviously acts as a chemical 
potential. 
\vskip 0.3cm

Despite the kinetic term of the vector field (we will discuss its role 
below), the model 
(\ref{eq:12}) is designed as a low-energy effective quark theory, with the 
quark loop momenta cut off at an energy scale $\Lambda $~\cite{la96}. 
We stress that the regularization procedure must not spoil the 
Euclidean O(4) invariance, that is, the Lorentz symmetry in Minkowski space. 
The magnitude of the cut-off $\Lambda $ is of the order of the gluonic 
energy scale. 
\vskip 0.3cm

In the chiral limit $m \rightarrow 0$, the Lagrangian 
$L$, (\ref{eq:13}), is invariant under chiral transformations. If the scalar
field acquires a non-vanishing vacuum expectation, i.e. 
$\langle \sigma \rangle \not=0$, the chiral symmetry is spontaneously 
broken (SSB). The beautiful concept of the spontaneous breakdown of 
chiral symmetry allows to interpret the light pseudo-scalar mesons as 
Goldstone bosons, and hence provides a natural and model-independent 
explanation of the particular role of the pions in the meson mass spectrum. 

\begin{figure}[t]
\centerline{ 
\epsfxsize=9cm
\epsffile{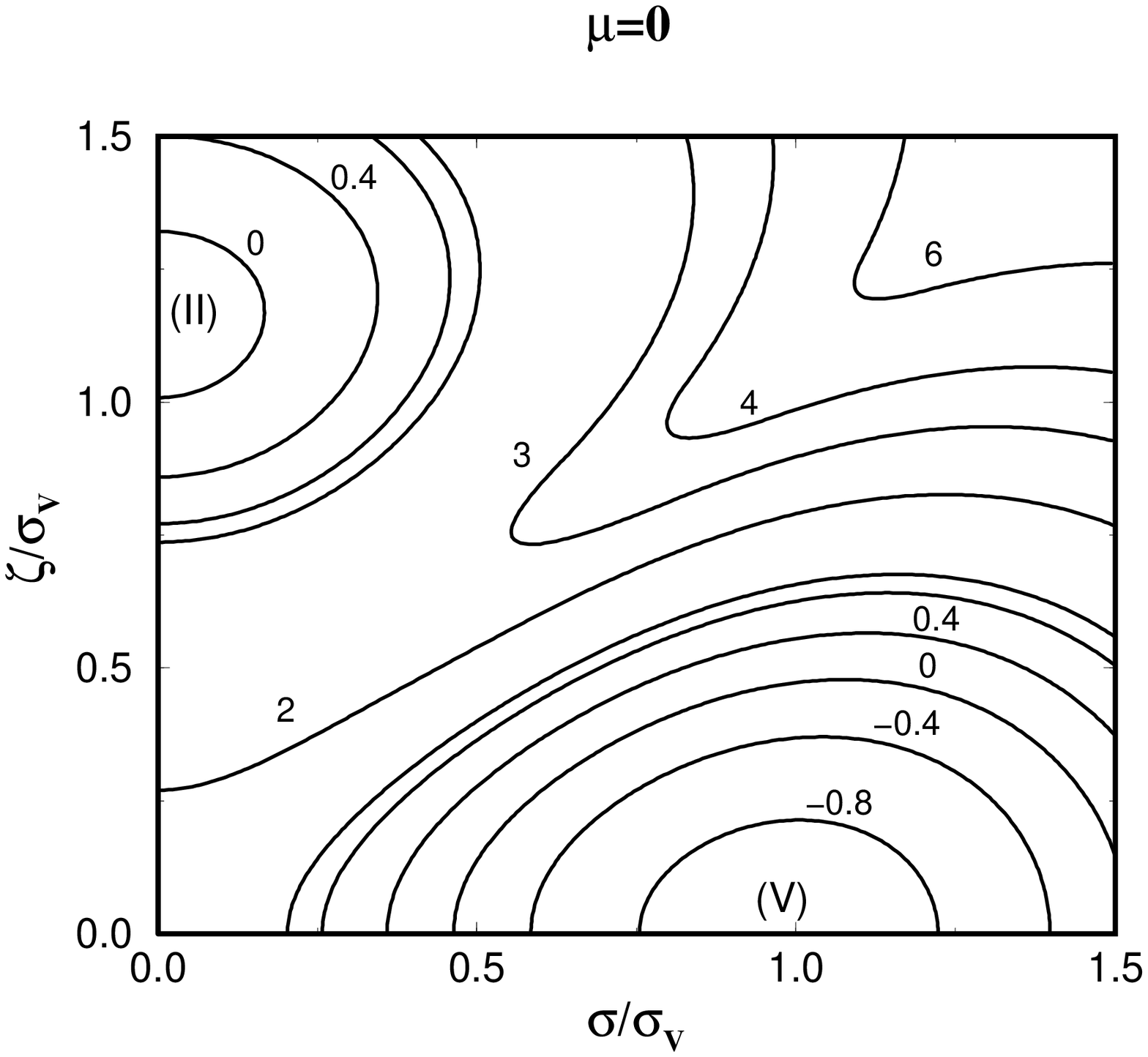} 
\epsfxsize=9cm
\epsffile{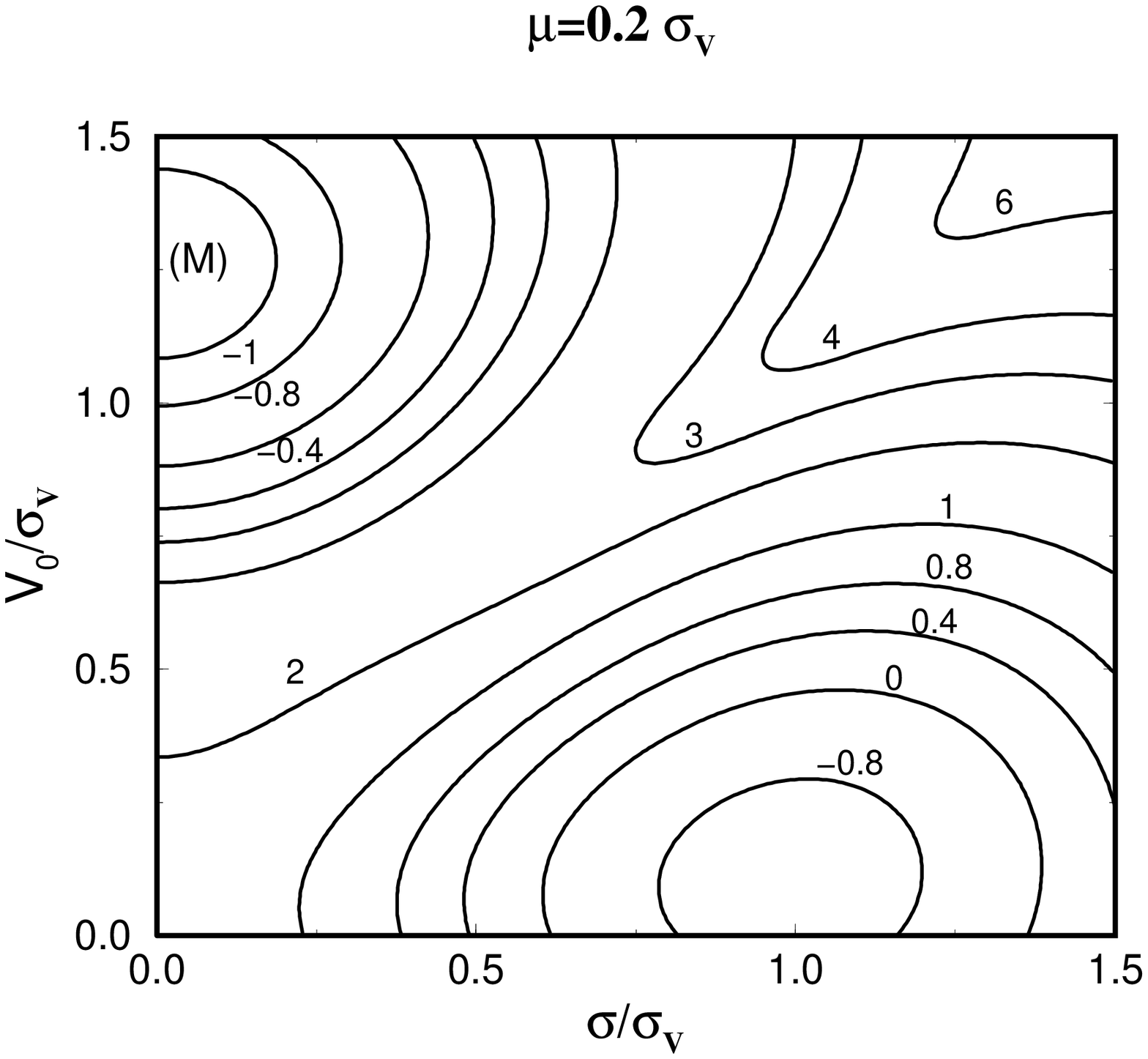} 
}
\vspace{-.8cm} 
\caption{ Lines of constant effective potential $U$ in the plane of $\sigma $ 
   and $\zeta ^2=V_\mu V_\mu $ for $\mu =0 $ (left picture) and in the plane 
   of $\sigma $ and $V_0$ for $\mu =0.2 \sigma _V $ (right picture). 
   $\sigma _V$ is the vacuum expectation value of $\sigma $. 
   The numbers provide the effective potential $U$ in units of 
   $\vert U_V \vert $, where $U_V$ is the vacuum value of the potential $U$. 
   $(V)$ and $(M)$ indicate the position of the vacuum state and of the 
   ground state in matter, respectively. } 
\label{fig:1} 
\end{figure} 
\vskip 0.3cm

\subsection{ Ground state properties } 

We will now show that the simple model (\ref{eq:12}) exhibits a 
vacuum phase $(\mu =0)$ where chiral symmetry is spontaneously broken (SSB)
and a $\mu $-induced transition to a phase where the (scalar) quark condensate 
vanishes and a spontaneous breakdown of Lorentz symmetry occurs (ISSB) 
on top of the explicit breaking. The convenient quantity by means of 
which this mechanism can be illustrated is the effective 
potential as function of the scalar and vector fields, i.e. 
$U(\sigma, V_\mu )$. The effective action $\Gamma $ is defined by a 
Legendre transform of the generating functional $\ln \, Z[s,j_\mu]$ with 
respect to the external sources $s(x)$ and $j_\mu (x)$, respectively, i.e. 
\bea 
\Gamma (\sigma , V_\mu ) &=& - \ln \, Z[s, j_\mu ] \; + \; 
\int d^4x \; \left[ s(x) \sigma (x) \, + \, j^\mu (x) V_\mu (x) 
\right] \; , \\ 
V_\mu (x) &=& \frac{ \delta \, \ln Z [s,j_\mu ] }{ \delta j^\mu (x) } 
\; , \hbo 
\sigma (x) \; = \; \frac{ \delta \, \ln Z [s,j_\mu ] }{ \delta s (x) }. 
\ena 
To 
leading order in the large $N$ expansion, it is sufficient to evaluate the 
functional integral (\ref{eq:12}) in a mean-field (stationary phase) 
approximation, since 
fluctuations around the mean-fields are suppressed by a factor $1/N$. 
A straightforward calculation yields 
\bea 
\frac{1}{N} \, \Gamma (\sigma, V_\mu ) &=& - \frac{1}{N} \Tr \, \ln 
\left\{ i \dslash \, - \, \sigma (x) \, + \, i \pi (x) \gamma _5 
\, + \, i \gamma ^\mu V_\mu (x) \right\} 
\label{eq:14a} \\ 
&+& \int d^4x \; \biggl\{ \frac{g_s}{2} \left[ (\sigma -m )^2 + \pi ^2 
\right] 
\nonumber \\
&+& \frac{1}{2} \left\{ V_\mu (x) \left( - \partial ^2 \delta _{\mu \nu } 
+ \partial _\mu \partial _\nu \right) V_\nu (x) \, + \, m_v^2 
\left[ (V_0 - \mu )^2 + V_k^2 \right] \right\} 
\biggr\} \; + \; {\cal O}(\frac{1}{N}) \;  
\nonumber 
\ena 
where the trace extends over Lorentz indices as well as over the Euclidean 
space-time. Note that a regularization which preserves the O(4) 
invariance is understood in (\ref{eq:14a}) in order to define the 
trace term. 
We then obtain the potential $U(\sigma, V_\mu )$ from the effective 
action by confining ourselves to constant classical fields. 
Assuming a vanishing mean field for the pionic field $\pi (x)$, 
we find for a sharp momentum cutoff (for details see appendix B) 
\bea 
\frac{1}{N} \, U(\sigma, V_\mu ) &=& - \frac{1}{8\pi ^3 } 
\int _{-1}^{+1} dx \; \sqrt{1- x^2} \int _{0}^{\Lambda ^2} du \; u \; 
\ln \left[ (u-\zeta ^2+\sigma ^2)^2 \, + \, 4 \zeta ^2 \, x ^2 \, u 
\right] 
\nonumber \\ 
&+& \frac{g_s}{2} \left[\sigma -m \right] ^2
\; + \; \frac{m^2_v}{2} \left[ (V_0 - \mu )^2 + V_k^2 \right] \; + \; 
{\cal O}(\frac{1}{N}) \;  
\label{eq:14} 
\ena 
where $\zeta ^2:=V_\mu V_\mu $. 
Due to O(4)--invariance, the effective potential $U$ depends only 
on the O(4)--invariant field combination $\zeta ^2 $ 
in the case $\mu =0$. Minima of the effective potential serve 
as possible candidates for the ground state. The global minimum, i.e. 
the state with the lowest vacuum energy density, represents 
the vacuum. The left-hand picture of Figure \ref{fig:1} 
shows the effective potential $U$ for $g_s = m^2_v = \Lambda ^2 / 8\pi^3  $. 
At zero chemical potential, the global minimum of $U$ is located 
at $(V)$ in Figure \ref{fig:1}. The corresponding vacuum properties are 
precisely characterized by (\ref{eq:1}). Chiral symmetry is spontaneously 
broken. In addition, the effective potential possesses a local minimum 
(as indicated by $(II)$ in Figure \ref{fig:1}). This minimum corresponds 
to a meta-stable state with the properties (\ref{eq:3}). 
\vskip 0.3cm

The picture changes drastically, if the chemical potential is increased. 
If the chemical potential exceeds a critical strength $\mu_c$, the global
minimum flips from the state (V) to the state (II) (see right hand side 
of figure 1). Since the chemical potential 
selects the zeroth component of the O(4)--invariant combination 
$V_\mu V_\mu $, the ground state in matter will be 
characterized by (\ref{eq:2}). This means that
in addition to a small explicit breaking, the O(4) (Lorentz) symmetry is  
spontaneously broken. Thus the model (\ref{eq:12}) exhibits the 
ISSB-mechanism discussed above. Finally, let us mention that a phase 
structure similar to the one of our toy model has been also found 
in phenomenologically successful effective nucleon-meson theories~\cite{x}. 
An analogous phenomenon occurs in the random-matrix study of the QCD phase
transition in the presence of chemical potential\cite{nowak}.

\subsection{ The ``almost'' Goldstone vector boson } 

In this subsection, we study small fluctuations $v_\mu (x)$ 
of the vector field $V_\mu (x)$ around its mean-field value $V^B_\mu $, 
i.e. $V_\mu (x) = V^B_\mu + v_\mu (x)$. We assume that a small 
chemical potential is sufficient to induce the transition from the state 
V to the state M and treat the influence of the small explicit 
breaking of the O(4) symmetry by the chemical potential $\mu $ 
as a perturbation. 
We will find that in this case the vector fields $v_{k=1 \ldots 3}(x)$ 
emerge as massless excitations from the Bethe-Salpeter equation, if 
$\mu $ goes to zero. 
For this purpose, we expand the effective action (\ref{eq:14a}) up to 
second order in the fields $v_\mu $, i.e. 
\bea 
\frac{1}{N} \, \Gamma ^{(2)} &=&  \frac{1}{2N} \, \Tr \left\{ 
\frac{i}{ i \dslash + i \Vslash _B - \sigma } \, \vslash 
\frac{i}{ i \dslash + i \Vslash _B - \sigma } \, \vslash \right\} 
\label{eq:20} \\ 
&+& \frac{1}{2} \int (p) \; v_k (p) \, \left[ (p^2 \delta _{kl } 
- p_k p_l ) \, + \, m_v^2 \right] \, v_l (-p) \; 
\nonumber 
\ena 
where $(p)$ is the shorthand for $d^4 p/(2\pi)^4$.
The explicit calculation of the trace term in (\ref{eq:20}) is left 
to appendix C. The final result can be written as 
\be 
\frac{1}{N} \, \Gamma ^{(2)} \; = \;  \frac{1}{2} \int (p) \; 
v_k(p) \; \Pi _{kl} (p) \; v_l(-p) \; . 
\label{eq:21} 
\en 
Mass eigenstates appear as solutions of the Bethe-Salpeter equation 
\be 
\Pi _{kl} (p_0^2 = - m^2_V, \vec{p}=0 ) \; v_l (-p) \; = \; 0 \; , 
\label{eq:22} 
\en 
where $m_V$ is the mass of the excitation. For simplicity, we 
here consider the Bethe-Salpeter equation for vanishing spatial 
momentum. It is also sufficient for our purposes 
to study this equation in a derivative 
expansion with respect to the meson momentum $p$\footnote{Note that 
the derivative expansion becomes exact for (massless) Goldstone 
bosons.}. 
Exploiting the gap equation for the mean field $V^B_\mu = (V_0,0,0,0)$, 
one finds (see appendix C) 
\be 
\Pi _{kl} (p_0^2, \vec{p}=0 ) \; = \; [ 1 - f(\sigma, V^B_0)] 
\, p^2 \; \delta _{kl} \; + \; \frac{ m_v^2 }{V_0} \, \mu \, 
\delta _{ik} \; . 
\label{eq:23} 
\en 
The main observation is that for $\mu =0 $ and $V_0 \not= 0$ the mass 
term drops out and that a massless excitation $(p^2=0)$ 
with the quantum numbers of a (non-relativistic) vector field 
occurs. For a small explicit breaking 
$\mu $ of the Lorentz symmetry, one can cast the Bethe-Salpeter equation 
(\ref{eq:22}) into $(p^2 = - m_V^2)$ 
\be 
[ 1 - f(\sigma, V^B_0)] \, m_V^2 \; = \; \frac{ m_v^2 }{V_0} \, \mu . 
\label{eq:23a} 
\en 
Using the gap equation (\ref{eq:c1}) (in appendix C), we find 
\be 
m_v^2 V_0 \; = \; \langle \bar{q} \gamma _0 q \rangle \; 
+ \; {\cal O}(\mu ) \; . 
\label{eq:23b} 
\en 
With this result, eq.(\ref{eq:23a}) can be cast into the form of 
eq.(\ref{eq:10}), i.e. 
\be 
f_V^2 \, m_V^2 \; = \; 2 \, \mu \, \langle \bar{q} \gamma _0 q \rangle \; 
+ \; {\cal O}(\mu ^2) \; , \hbo 
f_V \; = \; \sqrt{2} \, \sqrt{  1 - f(\sigma, V^B_0) } \; V_0 \; . 
\label{eq:23c} 
\en

\begin{figure}[t]
\parbox{7cm}{ 
\centerline{ 
\epsfxsize=7cm
\epsffile{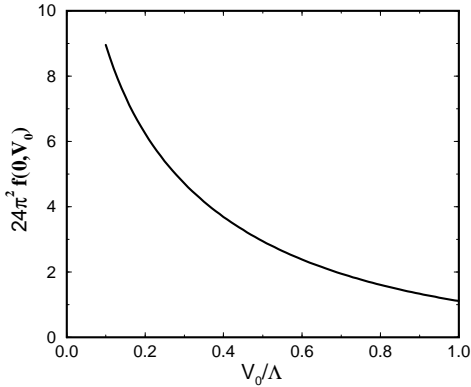} 
}
} \hspace{1cm}
\parbox{6cm}{ 
\caption{ The function $f(0,V^B)$ as function of $V_0$ in units of the 
   cutoff $\Lambda $. 
} }
\label{fig:2} 
\end{figure} 
If the first-order phase transition from the vacuum state (V) to the 
matter state (M) takes place, the constituent quark mass 
$\sigma $ drops to the value of the current mass. Therefore, we 
expect $\sigma \ll V_0$. Let us study the function $f(\sigma, V^B)$ 
for the case $\sigma =0$. The explicit calculation of this function is 
left to appendix C. The function $f$ is shown in Figure 2. 
It diverges logarithmically at the origin and decreases rapidly 
for large values of $V_0$. 
\vskip 0.3cm

It turns out that the function $f(\sigma, V^B_0)$ in (\ref{eq:23}) 
is always positive. The ``1'' in the square bracket stems from the bare 
kinetic term of the vector fields in (\ref{eq:13}), whereas $f$ is the 
contribution of the quark loop to the vector kinetic term. One observes 
that the quark-loop-induced kinetic term favors gradients in the 
vector field $v_k(x)$. If the parameter set is chosen such that 
$f(\sigma, V^B_0) < 1$, we can interpret the small amplitude fluctuations 
$v_k(x)$ as particle excitations in the usual way. 
In the case $f(\sigma, V^B_0) \ge 1 $, the small amplitude fluctuations 
exponentially grow, and the matter state would favor large gradients. 
This would probably lead to the formation of domain walls. 
In the present model, $f$ is always much less than $1$ in the parameter range 
of interest. We believe that $f>0$ is a generic feature, implying 
that quark loop contributions support instabilities. 
In contrast, we expect that the parameters $(\sigma, V_0)$ that would produce 
instability (i.e., leading to $f=1$) are highly model-dependent.

\vskip 0.3cm
 
\section{ The Particle Spectrum of the ISSB Scenario } 
\vskip 0.3cm 

The toy model considered above with one flavor of quarks gives ``almost''
Goldstone vectors of the $\omega$ meson quantum number. To be realistic,
we need at least two light flavors. Let us consider this case. 
For vanishing current mass and chemical potential, the quark sector 
exhibits chiral symmetry, i.e. 
\be 
SU_V (2) \times SU_A(2) \; , 
\label{eq:15a} 
\en 
and Lorentz invariance. In particular, the $SU_A(2)$ transformations 
relate scalar particles with pseudo-scalar particles, and transform 
vector current into axial-vector currents. The vacuum state is characterized 
by a spontaneous breakdown of the axial part of (\ref{eq:15a}). According to 
Goldstone's theorem, each generator of the spontaneously broken symmetry 
gives rise to a massless particle. In the case of QCD, the iso-triplet pions 
can be identified with the Goldstone bosons. However, 
the iso-triplet pions are not massless, but possess a mass which is small 
compared with the hadronic energy scale, since the chiral symmetry is 
also explicitly broken by small current masses. 
\vskip 0.3cm 

In the case of the ISSB scenario, the light particle content of the 
spectrum changes drastically. The crucial fact is the occurrence of the 
condensate (that is, in \break Minkowski space)
\be 
\langle \, \bar{q} \, \gamma _0 \, 1 \, q \, \rangle \; , 
\label{eq:15} 
\en 
where the unit operator in (\ref{eq:15}) indicates that the condensate is 
iso-scalar.  Lorentz symmetry is spontaneously broken over and above the 
explicit breaking via the chemical potential. 
On the other hand, the quark condensate $\langle \, \bar{q} q \, \rangle $ 
is proportional to the current quark mass and vanishes in the chiral 
limit. Since the condensate (\ref{eq:15}) is invariant under a chiral 
rotation of the quark fields, chiral symmetry is not spontaneously broken 
and the mass of the pions is not necessarily small. 
\vskip 0.3cm

The quantum numbers of the Goldstone vector bosons can be most 
easily seen by going to Euclidean space where the Lorentz group 
becomes the SO(4) group, which is equivalent to SU(2)$\times $SU(2). 
Expanding the anti-symmetric matrices $\omega _{\mu \nu }$ 
of the Euclidean (Lorentz) transformation (\ref{eq:a81}) in terms 
of the 't~Hooft symbols~\cite{hooft}  
$\eta ^i_{\mu \nu }$, $\bar{\eta }^i _{\mu \nu }$, 
which form a complete basis of self-dual and anti-self-dual matrices, 
\be 
\omega _{\mu \nu } \; = \; \theta _k \eta ^k_{\mu \nu } \, + \, 
\bar{\theta }_k \bar{\eta }^k_{\mu \nu } \; , 
\en 
the generators of Euclidean (Lorentz) transformations can be written as 
\be 
S(\Lambda ) \; = \; \exp \left[ -i \left( \theta _k \Sigma ^k_R 
\, + \, \bar{\theta } _k \Sigma ^k_L \right) \right] \; , 
\label{eq:x} 
\en 
where 
\be 
\Sigma ^k_R \; = \; \frac{1}{4} \eta ^k _{\mu \nu } \sigma ^{\mu \nu } \; , 
\hbo 
\Sigma ^k_L \; = \; \frac{1}{4} \bar{\eta } ^k _{\mu \nu } 
\sigma ^{\mu \nu } \; . 
\en 
In the direct product representation of the $\gamma $-matrices~\cite{itz}, 
one finds~\cite{hr91} 
\be 
\Sigma ^k_{R/L} \; = \; P_{R/L} \times \sigma ^k \; , \hbo 
P_{R/L} = \frac{1}{2} ( 1 \pm \gamma _5 ) \; . 
\label{eq:y} 
\en 
The matrices $P_{R/L}$ are the right and left handed projectors and 
$\sigma ^k $ are the familiar Pauli spin matrices. With (\ref{eq:y}) the 
Euclidean 
(Lorentz) transformation (\ref{eq:x}) has precisely the form of a chiral 
transformation with the iso-spin (or in general flavor) matrices 
replaced by the spin matrices. Defining $ 
\theta ^k_{V/A} = \frac{1}{2} ( \theta ^k \pm \bar{\theta }^k ) $, 
eq.~(\ref{eq:x}) becomes 
\be 
S(\Lambda ) \; = \; \exp \left[ -i \theta ^k_V \, (1 \times \sigma ^k) 
\, - \, i \bar{\theta } ^k_A (\gamma _5 \times \sigma ^k) \right] \; . 
\label{eq:z} 
\en 
This representation of the Euclidean (Lorentz) group corresponds to the 
coset decomposition 
\be 
SU_L(2) \times SU_R(2) \; = \; SU_V(2)\times 
\left[ SU_L(2) \times SU_R(2) / SU_V(2) \right] \; . 
\en 
It is the coset symmetry $SU_L(2) \times SU_R(2) / SU_V(2) $ 
which is spontaneously broken in the state (M) (\ref{eq:2}). 
Using the analogy between the Lorentz transformations (\ref{eq:z}) 
and a chiral transformation, it becomes clear that in the same way 
as the Goldstone bosons of spontaneous broken chiral symmetry carry 
iso-spin, the massless particles of spontaneously broken Lorentz symmetry 
carry spin and are hence vector particles. 
They are given by the spatial components of iso-scalar vectors 
$\omega _i := \bar{q} \, \gamma _i \, 1 \, q $. 
\vskip 0.3cm 

The iso-triplet $\rho $-mesons with a reduced mass in matter play a central 
role for the explanation of the dilepton enhancement in the CERES and 
HELIOS experiments in the approach of~\cite{ko96}. It is therefore 
interesting to look at the properties of the $\rho $-mesons in the ISSB 
scenario. First note that in the vacuum  the $\omega $-meson is 
somewhat heavier than 
the $\rho $-meson and that the overlap of the 
$\rho $--meson and $\omega $-meson wave-functions is non-zero~\cite{rom95}. 
At some large density, a light meson with the quantum numbers of the 
$\omega $-meson appears as an ``almost'' 
Goldstone boson. As density decreases from the critical density as the 
system expands, it can happen that the levels cross with  
the $\omega $ and $\rho $ becoming degenerate at the crossing point. 
In this case, 
the $\rho \omega $-mixing will become $50\%$ independently of the strength 
of the overlap matrix elements. 
\vskip 0.3cm

\goodbreak 
\section{ Discussions and Conclusions } 
\vskip 0.3cm 

We have shown that a meta-stable state with the properties (\ref{eq:3}) 
exists at vanishing chemical potential, if the effective low-energy quark 
interaction in the time component of the vector channel is attractive and 
strong enough. Using a simple effective quark model, we argued 
that this situation is 
generic for a wide range of parameter choices. 
We have suggested that the presence of this meta-stable state has 
important consequences 
at finite baryon density. If the chemical potential exceeds a critical 
value, a first-order phase transition from the vacuum phase 
to the former meta-stable state can take place. The scalar quark condensate 
vanishes and chiral symmetry is restored. 
In this phase, pions are no longer Goldstone bosons. 
In this new state of matter (state (M)), Lorentz symmetry is spontaneously 
broken (over and above the explicit breaking via the chemical potential), 
and light iso-scalar vector particles are excited. The light iso-scalar 
vector particles will dominantly couple to photons. 
\vskip 0.3cm

\begin{figure}[t]
\parbox{7cm}{ 
\centerline{ 
\epsfxsize=7cm
\epsffile{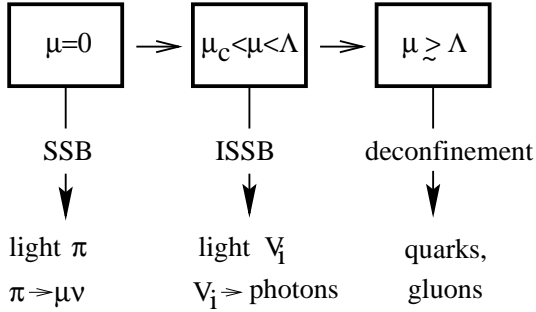} 
}
} \hspace{1cm}
\parbox{6cm}{ 
\caption{ The light particle content at several values of the chemical 
  potential. $\Lambda $ here is the fundamental energy scale of QCD. 
} }
\label{fig:3} 
\end{figure} 
If the density is further increased until it becomes large 
compared with the fundamental gluonic energy scale, one expects that 
the interaction between the quarks becomes weak due to asymptotic freedom. 
In this case, the iso-scalar condensate $\langle \bar{q} \, 
\gamma _0 \, 1 \, q \, \rangle $ loses the strong contribution from 
the interaction and will scale proportional to the chemical potential. 
This implies that the Goldstone mechanism no longer applies and that the 
iso-scalar vector mesons become heavy again before they dissolve in 
$q$-$\bar{q}$ pairs at very high density, where 
one expects a phase transition to the quark gluon plasma. 
The content of light particles of hadronic matter for several values 
of the density is illustrated in Figure 3. 
\vskip 0.3cm

What are possible implications of  the new state of matter at medium 
densities in heavy-ion collisions? Since the phase transition from the 
vacuum state to the new matter state is  first order, the matter state (M) 
will appear in bubbles in the standard matter phase where the mesons 
experience a small change of the vacuum properties. The experimental 
observation could be a superposition of the results of the standard ``hadronic 
cocktail'' and of the ISSB scenario. The quantitative outcome 
will depend on how much of the bubbles are nucleated. It is obvious that 
the contribution of the new matter state to the observables will be more 
pronounced in Pb-Au than in p-Au collisions. It would be interesting to see 
whether this scenario has any role in the dilepton enhancement
actually observed in the  CERES and HELIOS experiments.

\vspace{1cm} 
\centerline{\bf Acknowledgments: \hfill }
\vskip 0.3cm
We are grateful to Maciek Nowak for helpful comments on the manuscript. 
One of us (KL) thanks Herbert M\"uther for interesting informations 
on the $\rho $-meson condensation in nuclear matter, and 
Herbert Weigel for critical discussions. MR is grateful for the support of
IBERDROLA de Ciencia y Tecnologia of Spain and for the hospitality
of Vicente Vento in the theory group of Valencia.

\vspace{1cm} 

\appendix 
\section{ Notation and conventions } 

The metric tensor in Minkowski space is 
\be 
g_{\mu \nu } \; = \; \hbox{diag} ( 1 , -1 , -1 , -1 ) \; . 
\label{eq:a1} 
\en 
We define Euclidean tensors $T_{(E)}$ from the tensors in Minkowski 
space $T_{(M)}$ by 
\be 
T^{\mu _1 \ldots \mu _N }_{ (E) \phantom{\ldots \mu _N } \nu _1 
\ldots \nu _n } 
\; = \; (i)^r \, (-i)^s \; 
T^{\mu _1 \ldots \mu _N }_{ (M) \phantom{\ldots \mu _N } \nu _1 
\ldots \nu _n } \; , 
\label{eq:a2} 
\en 
where $r$ and $s$ are the numbers of zeros within $\{ \mu _1 \ldots 
\mu _N \}$ and $\{ \nu _1 \ldots \nu _n \} $, respectively. 
In particular, we have for the Euclidean time and the Euclidean 
metric 
\be 
x_{(E)} ^ 0 \; = \; i \, x_{(M)}^0 \; , \hbo 
g^{\mu \nu }_{(E)} \; = \; \hbox{diag} ( -1 ,-1 ,-1 ,-1 ) \; . 
\label{eq:a3} 
\en 
Covariant and contra-variant vectors in Euclidean space differ by an 
overall sign. For a consistent treatment of the symmetries, one is forced 
to consider the $\gamma ^\mu $ matrices as vectors. Therefore, one is 
naturally led to anti-hermitian Euclidean matrices via (\ref{eq:a2}), 
\be 
\gamma ^0 _{(E)} = i \gamma ^0 _{(M)} \; , \hbo 
\gamma ^k _{(E)} = \gamma ^k _{(M)} \; . 
\label{eq:a4} 
\en 
In particular, one finds 
\be 
\left( \gamma ^\mu _{(E)} \right) ^{\dagger } \; = \; 
- \; \gamma ^\mu _{(E)} \; , \hbo 
\{ \gamma ^\mu _{(E)} , \gamma ^\nu _{(E)} \} \; = \; 
2 g^{\mu \nu }_{(E)} \; = \; -2 \, \delta _{\mu \nu } \; . 
\label{eq:a5} 
\en 
The so-called Wick rotation is performed by considering the 
Euclidean tensors (\ref{eq:a2}) as real fields. 

\vskip 0.3cm
In addition, we define the square of an Euclidean 
vector field, e.g. $V_\mu $, by 
\be 
V^2 \; := \; V_\mu V_\mu \; = \; - V_\mu V^\mu \; . 
\label{eq:a51} 
\en 
This implies that $V^2$ is always a positive quantity (after the wick 
rotation to Euclidean space). 

\vskip 0.3cm
The Euclidean action $S_E$ is defined from the action $S_M$ in Minkowski 
space by 
\be 
\exp \{ i S_M \} \; = \; \exp \{ S_E \} \; . 
\label{eq:a52} 
\en 
Using (\ref{eq:a2}), it is obvious that the Euclidean Lagrangian $L_E$ 
is obtained from the Lagrangian $L_M$ in Minkowski space by replacing 
the fields in Minkowski space by Euclidean fields, i.e. 
$L_E = L_M$. 

\vskip 0.3cm
Let the tensor $\Lambda ^{\mu }_{\phantom{\mu } \nu } $ denote a 
Lorentz transformation in Minkowski space, i.e. 
\be 
\Lambda ^{\mu }_{\phantom{\mu } \alpha } \Lambda ^{\nu }_{\phantom{\nu } 
\beta } \, g^{\alpha \beta } \; = \; g^{\mu \nu } \; . 
\label{eq:a6} 
\en 
Using the definition (\ref{eq:a2}), one easily verifies that 
$\Lambda ^{\mu }_{(E) \, \nu } $ are elements of an O(4) group, i.e. 
$ \Lambda ^{T}_{(E)} \Lambda _{(E)} = 1$, which is the counterpart 
of the Lorentz group in Euclidean space.
\vskip 0.3cm

In order to define the Euclidean quark fields, we exploit the 
spinor transformation of the Euclidean quark field 
\bea 
q_{(E)}(x_E) \rightarrow q^\prime _{(E)}(x^\prime _E) &=& 
S(\Lambda _{(E)}) q_{(E)}(x_E) \; , \hbo 
x_E \rightarrow x^\prime _E \; = \; \Lambda x_E \; , 
\label{eq:a7} \\ 
S(\Lambda _{(E)}) \gamma ^\mu _{(E)} S^\dagger (\Lambda _{(E)}) 
&=& \left( \Lambda ^{-1} _{(E) } \right)^{\mu }_
{\phantom{\mu } 
\nu } \, \gamma ^\nu _{(E)} \; , 
\label{eq:a8} 
\ena 
where the matrices 
\be 
S(\Lambda _{(E)}) \; = \; \exp \{ - \frac{i}{4} \omega _{\mu \nu } 
\sigma ^{\mu \nu }_{(E)} \} 
\label{eq:a81} 
\en 
are unitary. It is obvious that one must interpret 
\be 
\bar{q}_{(E)} \; = \; q^\dagger _{(E)} 
\label{eq:a9} 
\en 
in order to ensure that e.g.~the quantity $\bar{q}_{(E)} \gamma ^\mu _{(E)} 
q_{(E)} $ transform as an Euclidean vector.
We suppress the index $E$ throughout the paper and mark tensors 
with an index $M$, if they are Minkowskian.

\section{ The gap equation } 

Let us calculate the trace term in the effective potential 
(\ref{eq:14a}) for constant entries and for $\pi (x)=0$. 
For this purpose, we write the trace as a sum over 
all eigenvalues $\lambda $ of the Euclidean Dirac operator. 
For constant fields $\sigma $ and $V_\mu $, it is convenient to calculate 
the eigenvalues in momentum space. For a fixed momentum, one finds 
\be 
\big( \kslash \; - \; \sigma \; + \; i V_\mu \gamma ^\mu \big) 
\; \psi \; = \; \lambda (k) \; \psi. 
\label{eq:g1} 
\en 
A direct calculation yields 
\be 
\lambda _\pm (k) \; = \; - \sigma \, \pm \, i \, \sqrt{ (k+iV)^2 } \; , 
\en 
where each eigenvalue is $2N$--fold degenerated. The trace term is 
therefore given by 
\be 
- 2 {\cal V} \int _{k \le \Lambda } (k) \; \ln \, \left( \lambda _+(k) 
\lambda _-(k) \right) \; = \; 
- 2 {\cal V} \int _{k \le \Lambda } (k) \; \ln \, \left[ 
\sigma ^2 + (k+iV)^2 \right] \; , 
\label{eq:g3} 
\en 
where ${\cal V}$ is the Euclidean space-time volume, and $\Lambda $ 
is the sharp O(4) invariant cutoff. 
$(k)$ is the shorthand for $d^4k /(2\pi )^4$. 
The integrand in (\ref{eq:g3}) 
is complex. We will, however, see that the imaginary part drops out, 
if we perform the integration over the momentum. For this purpose, 
we write (\ref{eq:g3}) as 
$$
- {\cal V} \int _{k \le \Lambda } (k) \; \ln \, \left[ 
\sigma ^2 + (k+iV)^2 \right] \; - \; 
{\cal V} \int _{q \le \Lambda } (q) \; \ln \, \left[ 
\sigma ^2 + (q-iV)^2 \right] \; = \; 
$$ 
$$ 
- {\cal V} \int _{k \le \Lambda } (k) \; \ln \, \left[ 
(k^2+\sigma ^2 -V^2) ^2 \, + \, 4 \, (k \cdot V)^2 \right] \; , 
$$ 
where we have performed a change of integration variables $q = - k$. 
We finally obtain 
\be 
- \frac{\cal V}{4 \pi ^2} \int _0^\pi \frac{d\alpha }{\pi } \; \sin ^2 \alpha \; 
\int _0^\Lambda dk \; k^3 \; \ln \left[ \left( k^2+\sigma ^2 -V^2 \right)^2 \, 
+ \, 4 k^2 V^2 \, \cos ^2 \alpha \right] \; . 
\label{eq:g4} 
\en 
This expression directly enters the effective potential $U(\sigma , 
V_\mu )$ in (\ref{eq:14}). Note that for $V^2>\sigma ^2$ the integrand 
in (\ref{eq:g4}) becomes singular. It is, however, easy to show 
that this singularity is integrable and that no imaginary part 
is present. The singularity occurs for $k^2=V^2-\sigma ^2$ in the angle 
integral. Using the principal-value prescription, this integral yields 
\be 
2 \lim _{\epsilon \to 0} \int _{\epsilon }^{+1} dx \; 
\sqrt{1-x^2} \, \log x^2 \; = \; 
- \pi \left( \ln 2 \, + \, \frac{1}{2} \right) \; . 
\en

\bigskip 
An extremum of the effective potential occurs, if the (constant) vector 
fields satisfy the gap equation 
\be 
- \frac{1}{N \, {\cal V}} \Tr \left\{ \frac{i}{ i \dslash \, 
- \, \sigma \, + \, 
i \Vslash } \; \gamma ^\mu \right\} \; + \; m_v^2 \, (V_\mu - \mu \, 
\delta _{\mu 0} ) \; = \; 0 \; . 
\label{eq:g41} 
\en 
Let us study the case without an explicit breaking of the O(4) 
symmetry, i.e. $\mu =0$. Since we use an O(4)-invariant regularization 
of the space-time trace in (\ref{eq:g41}), the measure of the momentum 
integration is O(4) invariant. Let $V^B_\mu $ denote a solution of 
the gap equation (\ref{eq:g41}) (with $\mu=0$). Using the 
property (\ref{eq:a8}), one easily shows that the rotated field 
$\Lambda _{\mu \nu } V^B_\nu $ is also a solution. In this case, 
we multiply eq.(\ref{eq:g41}) with $V^B_\mu $ and obtain a single 
equation to determine the length $V^B_\mu V^B_\mu $ of the vector field. 
For $\mu \not= 0$, it is easy to show that a solution of (\ref{eq:g41}) 
is provided by $V^B = (V_0,0,0,0)$.
\vskip 0.3cm

Without loss of information, we calculate 
\be 
- \frac{1}{N} 
\; \Tr \, \left\{ \frac{i}{ i \dslash \, - \, \sigma \, + \, i \Vslash } 
\gamma ^\mu \, \right\} \; V_\mu 
\label{eq:g5} 
\en 
for constant fields $V_\mu $. Introducing momentum  eigenstates and 
performing the trace over Dirac indices yield 
\be 
- 4i {\cal V} \int _{k \le \Lambda } (k) \frac{ k \cdot V \, + \, i V^2 }{ 
(k+iV)^2 + \sigma ^2 } \; , 
\label{eq:g6} 
\en 
where $k \cdot V := k_\mu V_\mu = - k_\mu V^\mu $. 
Introducing polar coordinates where $\alpha $ denotes the angle between 
$k$ and $V$, one obtains 
$$ 
- \frac{i{\cal V}}{\pi ^3} \int _0^\pi d\alpha \; \sin ^2 \alpha \; 
\int _0^\Lambda dk \; k^3 \; \frac{ k V \, \cos \alpha \, + \, i V^2}{ 
k^2+\sigma ^2 -V^2 \, + \, 2 i k V \, \cos \alpha } \; = \; 
$$ 
\be 
- \frac{ V^2 }{ \pi ^3} {\cal V} \int _0 ^\Lambda dk \; k^3 \; 
\int _{-1}^{+1} dx \; \sqrt{1 - x^2} \; 
\frac{ 2 k^2 x^2 - k^2 -\sigma ^2 + V^2 }{ (k^2 + \sigma ^2 - V^2)^2 
\, + \, 4 k^2 V^2 \, x^2 } \; . 
\label{eq:g7} 
\en 
This expression of course agrees with the result which is obtained by 
taking the derivative of (\ref{eq:g4}) with respect to $V_\mu $ 
and multiplying with $V_\mu $.

\section{ The Bethe-Salpeter equation } 

In order to observe the cancelation of the mass term for the 
``almost'' Goldstone vectors, we need a certain relation which is 
satisfied for any solution $V^B_\mu $ of the gap equation 
(\ref{eq:g41}) with $\mu =0$. Our first task in this section is to 
derive this relation. 

For this purpose, we note that the trace term in (\ref{eq:g41}) transforms 
as an O(4) vector, i.e. 
\be 
B_\mu [V^B] \; = \; 
- 4i \, \int (k) \; \frac{ k_\mu + i V^B_\mu }{ (k+iV^B)^2 + \sigma ^2 } 
\; . 
\label{eq:c1} 
\en 
This is true, because we use an O(4) invariant regularization of the momentum 
integration. It is obvious that 
\be 
B _\mu [\Lambda V^B] \; = \; \Lambda _{\mu \nu } \, B_\nu [V^B] \; , 
\hbo 
\Lambda _{\mu \nu } \; = \; \exp \left\{ \theta ^a \eta ^a 
\right\} _{\mu \nu } \; . 
\label{eq:c2} 
\en 
The matrices 
$\eta ^a _{\mu \nu }$ are three of 't~Hooft's antisymmetric 
matrices, which serve as three out of six generators of the 
$O(4)$ transformation. Note that eq.(\ref{eq:c2}) is satisfied for 
any choice of the angles $\theta ^a$. Taking the derivative of this 
equation with respect to $\theta ^a$ yields the identity 
\be 
\eta ^a_{\mu \nu } B_\nu [V^B] \; = \; 
-4i \int (k) \; \frac{ i \eta ^a_{\mu \nu } V^B_\nu }{ (k+iV^B)^2 
+ \sigma ^2} \, + \, 4i \int (k) \; \frac{ (k+iV^B)_\mu \; 
2i \, (k_\alpha \eta ^a_{\alpha \beta } V^B_\beta ) }{ [ (k+iV^B)^2 
+ \sigma ^2]^2 } \;  . 
\label{eq:c3} 
\en 
If we specialize to $V^B_\mu = ( V_0, 0,0,0)$, use $\eta ^a_{l0} 
\; = \; \delta ^{al}$ and employ the gap equation, i.e. 
$B_\mu  = - m_v^2 (V_\mu - \mu \delta _{\mu 3})$, we finally obtain the 
desired equation ($V_0 \not=0 $), i.e. 
\be 
4 \int (k) \left[ \frac{ \delta ^{al} }{ (k+iV^B)^2 +\sigma ^2 } 
\, - \, \frac{ 2 k_l k_a }{ [(k+iV^B)^2 +\sigma ^2]^2 } 
\right] \; + \; m_v^2 \, \delta ^{al} \; = \; \frac{ m_v^2 }{V_0} \, \mu 
\, \delta _{al} \; . 
\label{eq:c4} 
\en 
\vskip 0.3cm

In the following, we evaluate the trace term in (\ref{eq:20}), 
which gives rise to the Bethe-Salpeter equation. 
Rewriting the trace as a sum over momentum eigenstates and 
inserting a complete set of these eigenstates, the trace 
term can be written as 
\be 
- \frac{1}{2} 
\int (p) \; v_\mu (p) v_\nu (-p) \; \int (k) \; 
\tr _D \left\{ \frac{ \kslash + \frac{ \pslash }{2} + i \Vslash ^B 
+ \sigma }{ \left( k + \frac{p}{2} + i V^B \right)^2 + \sigma ^2 } 
\gamma _\mu 
\frac{ \kslash - \frac{ \pslash }{2} + i \Vslash ^B 
+ \sigma }{ \left( k - \frac{p}{2} + i V^B \right)^2 + \sigma ^2 } 
\gamma _\nu \right\} \; , 
\label{eq:c5} 
\en 
where the trace $\tr _D$ extends over Dirac indices only. 
Performing the Dirac trace and introducing a Feynman integral, 
the polarization tensor in (\ref{eq:21}) is given by 
\bea 
\Pi_{ik}(p) &=& (p^2 \delta _{ik} - p_i p_k) \, + \, m_v^2 \, \delta _{ik} 
\label{eq:c6} \\ 
&-& 4 \int _0^1 dx \; \int (q) \; \frac{ 
2 q_i q_k - 2x(1-x) p_i p_k - \delta _{ik} \left[ (q + 
\frac{1-2x}{2} p + i V^B )^2 - \frac{p^2}{4} + \sigma ^2 
\right] }{ 
\left[ (q + iV^B)^2 + x(1-x) p^2 + \sigma ^2 \right]^2 } \; . 
\nonumber 
\ena 
In order to show that the vector fields are massless 
excitations for $\mu =0 $, we study the polarization tensor (\ref{eq:c6}) 
at zero momentum, i.e. 
\bea 
\Pi _{ik}(0) &=&  m_v^2 \delta _{ik} \, - \, 
4 \int (q) \; \frac{ 2 q_i q_k - \delta _{ik} \left( 
(q+iV^B)^2 +\sigma ^2 \right) }{ 
\left[ (q + iV^B)^2 + \sigma ^2 \right]^2 } 
\label{eq:c7} \\ 
&=& \frac{ m_v^2 }{V_0} \, \mu \, \delta _{ik} \; , 
\ena 
where we have used eq.(\ref{eq:c4}). 
The crucial observation is that the expression (\ref{eq:c7}) 
vanishes for $\mu =0 $ (and $V_0 \not= 0$).
\vskip 0.3cm

If we are interested in the polarization tensor close to the 
mass shell, it is sufficient to study the Bethe-Salpeter equation 
in a derivative expansion. It is straightforward to 
extract the order ${\cal O}(p^2)$ from (\ref{eq:c6}). 
One finally finds 
\be 
\Pi _{ik} (p_0,\vec{p}=0) \; = \; \delta _{ik} \, 
[1 - f(\sigma , V^B) ] \; p^2 \; + \; \frac{ m_v^2 }{V_0} \, \mu \, 
\delta _{ik} \; + \;  {\cal O}(p^4) 
\label{eq:c8} 
\en 
where 
\be 
f(\sigma , V^B) \; = \; \frac{2}{3} \int (q) \; 
\left\{ \frac{3}{ \left[ (q + iV^B)^2 + \sigma ^2 \right]^2 } 
\; - \; \frac{4}{3} \frac{\vec{q}^2 }{ 
\left[ (q + iV^B)^2 + \sigma ^2 \right]^3 } \right\}. 
\label{eq:c9} 
\en 

Let us study the interesting case $\sigma =0$, $V_0 \not= 0$. 
For this purpose, we introduce polar-coordinates for the momentum 
integration in (\ref{eq:c9}). The angle integration is tedious due 
to the complex functions. One must distinguish the cases $q >V_0 $ and 
$q < V_0 $. One finally obtains 
\be 
f(\sigma =0 , V^B) \; = \; \frac{1}{ 12 \pi ^2} \left\{ 
\int _0^{V_0} dq \; q^3 \; \frac{ q^2 + 3 V_0^2 }{ V_0^4 (q^2 + 
V_0^2) } \; + \; \int _{V_0}^ \Lambda dq \; q \; 
\frac{ 2 }{ q^2 + V_0^2 } \right\} 
\label{eq:c10} 
\en 
for $V_0 \le \Lambda $, and 
\be 
f(\sigma =0 , V^B) \; = \; \frac{1}{ 12 \pi ^2} 
\int _0^\Lambda dq \; q^3 \; \frac{ q^2 + 3 V_0^2 }{ V_0^4 (q^2 + 
V_0^2) } 
\label{eq:c11} 
\en 
for $V_0 > \Lambda $. The final momentum integration is straightforward. 
We obtain 
\bea 
f(\sigma =0 , V^B) &=& \frac{1}{24 \pi ^2} \left\{ 
\frac{5}{2} - 4 \ln 2 + 2 \ln \left( \frac{ \Lambda ^2 }{ V_0^2 } +1 
\right) \right\} \; , \hbox to 1cm {\hfil for \hfil } 
V_0 < \Lambda 
\label{eq:c12} \\ 
f(\sigma =0 , V^B) &=& \frac{1}{24 \pi ^2} \left\{ 
\frac{2 \Lambda ^2}{V_0^2} + \frac{\Lambda ^4}{2V_0^4} 
 - 2 \ln \left( \frac{ \Lambda ^2 }{ V_0^2 } +1 
\right) \right\} \; , \hbox to 1cm {\hfil for \hfil } 
V_0 > \Lambda \; . 
\label{eq:c13} 
\ena 
The final result depends logarithmically on the cutoff $\Lambda $ 
as expected from a naive power counting. We have numerically 
investigated the function $f(\sigma , V^B)$ for several values 
of $\sigma $. We find  that generically $f(\sigma , V^B) > 0$.

\end{document}